\documentclass{iopart}

\usepackage{iopams}  
\usepackage{setstack}
\usepackage{graphicx}
\usepackage{cite}
\usepackage{enumerate}

\usepackage{xcolor}
\usepackage{soul,ulem,amssymb}

\newcommand{\bd}{\begin{displaymath}}
\newcommand{\ed}{\end{displaymath}}
\newcommand{\be}{\begin{equation}}
\newcommand{\ee}{\end{equation}}
\newcommand{\ba}{\begin{eqnarray}}
\newcommand{\ea}{\end{eqnarray}}

\begin{document}

\title[]{Analysis of the gradual transition from the near to the far field in
single-slit diffraction}

\author{Almudena Garc{\'\i}a-S\'anchez and \'Angel S. Sanz}

\address{Department of Optics, Faculty of Physical Sciences\\
Universidad Complutense de Madrid\\
Pza.\ Ciencias 1, Ciudad Universitaria E-28040 Madrid, Spain}

\ead{\mailto{a.s.sanz@fis.ucm.es}}

\vspace{10pt}

\begin{abstract}
In Optics it is common to split up the formal analysis of diffraction according
to two convenient approximations, in the near and far
fields (also known as the Fresnel and Fraunhofer regimes,
respectively).
Within this scenario, geometrical optics, the optics describing
the light phenomena observable in our everyday life, is introduced as the
short-wavelength limit of near-field phenomena, assuming that the typical
size of the aperture (or obstacle) that light is incident on is much larger than
the light wavelength.
With the purpose to provide an alternative view on how geometrical optics fits
within the context of the diffraction theory, particularly how it emerges, the
transition from the near to the far field is revisited here both analytically
and numerically.
Accordingly, first this transition is investigated in the case of Gaussian beam
diffraction, since its full analyticity paves the way for a better understanding of the
paradigmatic (and typical) case of diffraction by sharp-edged single slits.
This latter case is then tackled both analytically, by means of some insightful
approximations and guesses, and numerically.
As it is shown, this analysis makes explicit the influence of the
various parameters involved in diffraction processes, such as the typical size of the input
(diffracted) wave or its wavelength, or the distance between the input and output planes.
Moreover, analytical expressions have been determined for the critical
turnover value of the slit width that separates typical Fraunhofer diffraction regimes
from the behaviors eventually leading to the geometrical optics limit, finding a good
agreement with both numerically simulated results and experimental data extracted from
the literature.
\end{abstract}

%
%
%
%
%


\section{Introduction}
\label{sec1}

Due to the wave nature of light, and more specifically the phenomenon of diffraction, the
intensity distribution observed behind a sharp opening is not homogeneous.
Consider the paradigmatic case of a very long rectangular slit \cite{bornwolf-bk,hecht-bk:2002}.
As is well known \cite{hecht-bk:2002}, the intensity distribution generated by this slit exhibits
a series of alternating bright and dark light fringes aligned parallel to the slit.
Accordingly, when observed along the transverse direction (perpendicular to the fringes), the profile of this
intensity distribution is described by a succession of maxima and minima, with the maxima decreasing in intensity
from the center of the pattern to the sides.
This phenomenon is observable whenever the wavelength of the incident light is comparable with the dimensions
of the slit (in this case, with respect to its width along the transverse direction).
Furthermore, it is also known \cite{elmore-heald-bk:1969} that, as the projection screen
where the intensity is observed moves further away from the slit, the intensity distribution
gradually changes from a highly oscillatory pattern, at distances relatively close
to the slit, to a stationary, scale invariant distribution, far enough from the slit.
At the experimental level, a very detailed account on this continuous transition from the
near field (Fresnel regime) to the far field (Fraunhofer regime) was reported by Harris
{\it et al.}~\cite{harris:JOSA:1969} in the late 1960s.
In particular, by considering slits with different width (which, in turn, leads to a gradual
change of the distance along the Cornu spiral), these authors showed how the diffraction pattern describing the
irradiance distribution smoothly changes from a typical Fraunhofer single-slit pattern to something that closely
resembles (neglecting highly oscillatory features) the light distribution right behind a sharp-edge opening.
As expected, these results were in agreement with numerical simulations carried out alongside.

The different oscillatory behavior observed in the Fresnel and Fraunhofer regimes arises
from the dependence of phase contributions in each case on the transverse coordinate
in relation to the distance between the slit and the projection screen \cite{bornwolf-bk}.
In the Fraunhofer regime, the linear dependence on this coordinate turns into a constant
aspect ratio in relative terms, which explains the scale-invariance of the intensity
pattern.
On the contrary, in the Fresnel regime, a quadratic dependence produces fast phase variations
from a distance to the next one, which translates, in turn, into a rapidly oscillatory
pattern.
As it has been mentioned above concerning the experiment reported in \cite{harris:JOSA:1969},
this pattern resembles the sharp shadow produced in the domain of geometrical optics, where the presence of
the slit has no effect on the incident collimated ray bundle, other than preventing the passage of rays
beyond the slit boundaries.
This effect becomes more apparent as the observation distance becomes closer and closer
to the slit.

As it is shown in \cite{panuski:PhysTeacher:2016}, the above situation naturally
leads to the question on how the transition from wave optics (represented by
Fraunhofer diffraction) to geometrical optics (somehow related to the first stages of a
Fresnel regime) takes place, that is, whether there is a distinctive trait in
such a transition which might help us to uniquely discriminate when we are in each regime
(other than, of course, the semiqualitative, well-known Rayleigh criterion).
Thus, following standard wave optics arguments \cite{elmore-heald-bk:1969}, we know that
at a fixed distance, $z$, from the projection screen (and for monochromatic light), the width of the
principal Fraunhofer diffraction maximum produced by a single slit shows a dependence with
the inverse of the slit width, henceforth denoted by $a$.
The width of this maximum thus falls as $a$ increases.
However, since $z$ is kept constant, such increase also implies that the Fraunhofer condition
is gradually lost; the Fresnel regime starts playing a role, making such a fall with $a$ to be no longer valid.
There is not a precise analytical way to determine the width of the main intensity maximum in
the Fresnel regime in relation to $a$, because of the fast oscillations that the intensity
distribution undergoes even with slight changes in $a$.
Nonetheless, to some extent it roughly resembles a distribution more typical of geometrical
optics, as mentioned above, with the shadow boundaries being determined by $a$.
In other words, it is reasonable to assume that, for large enough values of $a$, the width of
the main intensity maximum should increase linearly with $a$.

If those two arguments for the far and near fields must be satisfied, clearly the
intensity distribution should present a turnover (minimum) as $a$ is gradually increased,
which is precisely the result experimentally reproduced in \cite{panuski:PhysTeacher:2016}.
More recently, further numerical analyses have shown \cite{davidovic:PhysTeacher:2019} the
good agreement between those experimental data and the simple expression provided by the
scalar theory of diffraction for single-slit diffraction in the paraxial approximation
\cite{hecht-bk:2002,elmore-heald-bk:1969}.
Traditionally, this transition from the Fresnel regime to the Fraunhofer one is explained
by means of the Fresnel linear-zone model and the concept of
Cornu spiral \cite{elmore-heald-bk:1969}.
This methodology has been rather convenient from an algebraic point of view to understand
the diffraction phenomenon in the Fresnel regime, at least, when one tries to skip in as
much as possible numerical computations.
However, from an intuitive point of view, simpler analytical comparative models can help us
to better understand this transition, particularly taking into account that this is a
general trend, independent of the specific transmission properties of the opening
(although they might influence other aspects, such as the diffraction expansion rate or
the overall profile displayed by the intensity distribution).

With the purpose to provide an alternative understanding of such a transition and,
in particular, the appearance of the turnover as the slit width increases,
here we report on both analytical and numerical investigations of diffraction with two different
types of slits.
More specifically, first we present an analytical study and discussion of the phenomenon
in the case of Gaussian beam diffraction (which somehow mimics the behavior of a slit with a
Gaussian transmission function).
The simplicity of this fully analytical problem provides us with an intuitive first approach
to the issue, where it is readily seen that any diffraction process can be split up into
three different regimes depending on the expansion rate displayed by the diffracted beam,
which shall be denoted as the Huygens, Fresnel, and Fraunhofer regimes following
the analysis presented earlier on in \cite{davidovic:PhysTeacher:2019}.
The outcomes from this analysis are then applied to the case of the well-known sharp-edged
single-slit diffraction, which is numerically investigated in terms of the quantity also considered in
\cite{panuski:PhysTeacher:2016},
namely the full width at the 20\% of the principal maximum (FW02M) in order to compare
the results from our simulations with the experimental data reported by these authors.
It is seen that, when proceeding in a systematic way, while the dependence of the FW02M
on $a$ within the Fraunhofer regime is smooth, after crossing the turnover, a staircase
structure is seen to characterize the Fresnel regime, with shorter and shorter steps as $a$
increases and the system gets into the Huygens (geometrical) regime.

In both cases, the analytical expressions and numerical simulations
	considered show evidence that the geometrical optics regime is always characterized by
	a linear increase of the extension of the irradiance distribution, which is independent
	of the wavelength or the distance between input and output planes (for long distances),
	and regardless of the profile displayed by the diffracted field amplitude.
	Interestingly, this behavior arises after the typical size of the diffracted beam (or the
	slit width, in the case of sharp-edged slits) has overcome a critical turnover value,
	which is here analytically determined and compared with numerically simulated results
	as well as with experimental data extracted from the literature.
	Finally, the present analysis has also unveiled staircase structures in the near field, which seem to characterize
the trend towards the geometrical optics limit in this standardized
	diffraction problem, in sharp contrast to the smooth behaviors characterizing the far field.

The work is organized as follows.
The analysis and discussion of Gaussian-slit diffraction is presented in Sec.~\ref{sec2}.
Section~\ref{sec3} is devoted to diffraction by a very long rectangular slit, first
introducing a theoretical analysis and then discussing a series of experimental
results obtained with a simple arrangement (which is also described).
These results are also compared with the experimental data reported earlier on by Panuski and
Mungan \cite{panuski:PhysTeacher:2016}, finding a good agreement at the level of resolution of the experiment.
To conclude, the main findings observed here and some remarks connected to their
extrapolation to matter waves (where they should also be observable) are summarized
in Sec.~\ref{sec4}.


\section{Gaussian-slit diffraction}
\label{sec2}

It is known \cite{sanz-bk-1,sanz:ApplSci:2020} that the Helmholtz equation in paraxial form is
isomorphic to the time-dependent Schr\"odinger equation, with the longitudinal coordinate playing
the role of the evolution parameter (time in the of Schr\"odinger's equation).
Accordingly, a Gaussian beam displays the same dispersion along the transverse direction
than a quantum Gaussian wave packet does along time.
Actually, both solutions can be exchanged if the factor $\hbar t/m$ that rules the behavior
of a quantum Gaussian wave packet is substituted by a factor $z/k$ (or, equivalently,
$\lambda z/2\pi$, with $k = 2\pi/\lambda$), where $z$ denotes the longitudinal
coordinate along which the optical beam propagates \cite{sanz:ApplSci:2020}.
A paradigmatic example of Gaussian diffraction is that of a single-mode laser beam released
in free space.

Let us thus consider the electric field associated with such a laser beam \cite{sanz:ApplSci:2020},
\be
 {\bf E}({\bf r},z) = E_0\ \frac{w_0}{w_z}\ e^{-r^2/w_z^2 - i(kz + kr^2/2R_z) + i\varphi_z}
 \hat{\bf r} ,
 \label{Efield}
\ee
where ${\bf r}=(x,y)$ accounts for the radial (transverse) vector coordinate, $w_0$ is the beam
waist, and the other parameters are functions of the longitudinal $z$-coordinate, which accounts for the
distance between the output (observation) and input (launch) planes (the latter is taken as
the origin of the reference system).
In Eq.~(\ref{Efield}) there are several $z$-dependent functions, namely, the width of the beam at the output plane,
\be
 w_z = w_0 \sqrt{1 + \frac{z^2}{z_R^2}} ,
 \label{widthz}
\ee
the radius of curvature of the beam at such a plane,
\be
 R_z = z \left(1 + \frac{z_R^2}{z^2}\right) ,
\ee
and the Gouy phase,
\be
 \varphi_z = \left( \tan \right)^{-1} \left( \frac{z}{z_R} \right) ,
 \label{gouy}
\ee
all of them given in terms of the so-called Rayleigh range or distance,
\be
 z_R = \frac{k w_0^2}{2} = \frac{\pi w_0^2}{\lambda} .
 \label{rayleigh}
\ee
Accordingly, the irradiance or intensity distribution at the ouput plane $z$ is
\be
 I(r,z) = \frac{1}{2} \sqrt{\frac{\epsilon_0}{\mu_0}}
  \left\vert {\bf E}({\bf r},z) \right\vert^2
 = \frac{1}{2} \sqrt{\frac{\epsilon_0}{\mu_0}}\ \frac{w_0^2}{w_z^2} e^{-2r^2/w_z^2} ,
\ee
with $r = \sqrt{x^2 + y^2}$.
In this analysis the interest relies on the changes undergone by the intensity distribution
as the output plane, so we will consider relative intensities, i.e.,
\be
 I_{\rm rel} (r,z) = \frac{I(r,z)}{I(r=0,z)} = e^{-2r^2/w_z^2} .
 \label{relintens}
\ee

Since the intensity (\ref{relintens}) only depends on $z$ through $w_z$, from now on we shall only
focus on discussing its behavior in term of this parameter.
Thus, by inspecting (\ref{widthz}), three different stages or regimes in the propagation of
the beam along the $z$-direction can be distinguished, each one characterized by very specific
features:
\begin{itemize}
\item The geometrical or Huygens regime, for $z_R \ggg z$, where the beam does not
exhibit any remarkable dispersion, but remains essentially with the same width, since
$w_z \approx w_0$:
\be
 I_{\rm rel} (r,z) \approx e^{-2r^2/w_0^2} .
\ee
In this regime, diffraction effects are negligible (Huygens principle applies, strictly speaking)
and, therefore, the wavefronts are plane parallel ($R_z \approx z_R^2/z \to \infty$).
Thus, although diffraction-free propagation is typically associated with the size of the
wavelength, this case shows that the phenomenon also appears at the very early stages of the
field propagation, in close and direct analogy to the Ehrenfest regime in quantum mechanics
\cite{sanz:AJP:2012}.
Also note that the Gouy phase (\ref{gouy}) can be neglected ($\varphi_z \approx 0$) and
hence it has no influence on the beam propagation.

\item The far field or Fraunhofer regime, the opposite limit, for $z_r \ll z$,
where the beam width increases linearly with the $z$-coordinate,
$w_z \approx w_0 z/z_R = \lambda z/\pi w_0^2$, and the wavefronts are nearly spherical, with
their radius increasing linearly with $z$ ($R_z \approx z$).
In this case, the relative intensity distribution remains invariant due to its dependence
on the ratio $x/z$, i.e.,
\be
 I_{\rm rel} (r,z) \approx e^{-2r^2 z_R^2/w_0^2 z^2} = e^{- (2 \pi w_0^2/\lambda^2) (r/z)^2} .
\ee
In other words, the distance between two points in the distribution increases at the same
rate independently of the distance $z$ at which the distribution is observed.
As it is inferred from (\ref{gouy}), also here the phase remains essentially constant along
the field propagation, being $\varphi_z \approx \pi/2$.

\item The near field or Fresnel regime, at intermediate ranges, where the curvature of the
wavefronts is undefined and the Gouy phase is a varying function of the $z$-coordinate.
Specifically, in this regime the beam dispersion undergoes a rather fast boost, which
provokes an also fast increase of the beam width.
Thus, as $z$ increases, this boost phase leads the beam from an almost dispersion-free behavior to another
one characterized by a slower asymptotic linear increase.
This regime is still within the range of small $z$ ($z_r \gg z$), so that
\be
 w_z \approx w_0 \left(1 + \frac{z^2}{z_R^2}\right) .
\ee
This quadratic dependence on the longitudinal coordinate resembles an acceleration,
which depends quadratically on the time variable.
Nonetheless, it is seen from (\ref{rayleigh}) that the smaller the waist and the larger the
wavelength, the shorter the Rayleigh range, which indicates that the boost phase
takes place at shorter distances from the input plane.
It can also be shown that local phase variations in (\ref{Efield}) are actually ruled by
a quadratic dependence on the transverse coordinates $(x,y)$, unlike the Fraunhofer regime, where
such dependence is linear.
\end{itemize}
Of course, the transition from negligible to large $z$ is continuous, and hence there are
intermediate regimes.
Yet, the three regimes described above can always be clearly identified in any
diffraction process regardless of the initial shape of the beam, thus becoming distinctive spatial
traits of diffraction.
Therefore, although they have been determined on the basis of a Gaussian-slit diffraction process,
because of its full analyticity (and hence because of analytical convenience), the same conclusions
apply to the standard case of diffraction of a plane wave by a sharp-edged slit,
as it will be seen next, in Sec.~\ref{sec3}.
These considerations are not only in agreement with the analysis and discussion presented
in \cite{davidovic:PhysTeacher:2019}, in the context of the current experiment, but there
is also a close connection with the spreading of Gaussian wave packets in quantum mechanics
\cite{sanz:AJP:2012,sanz:JPhysConfSer:2012}.

The general features discussed above help us to understand how the width of a diffracted
beam evolves as it propagates.
Leaving aside the position of the output plane, the width of any beam essentially depends
on both the wavelength $\lambda$ of the incident radiation and the input width (the waist
$w_0$ at the input plane $z=0$, in the case of the Gaussian beam), which can be related,
in turn, to the size of the diffracting opening.
These quantities have an influence on the beam diffraction, measurable in terms of the size
of the beam at a given $z$ and the relative intensity.
In order to quantify the diffractive effect on the beam, we have considered the above-mentioned
measure of the full width at the 20\% of the maximum, FW02M, which depends on
both physical parameters, as it is seen below.
In particular, in the case of the Gaussian beam, we have
\be
 {\rm FW02M}(z;w_0,\lambda) = \sqrt{2 \ln 5}\ w_z
  = \sqrt{2 \ln 5}\ w_0\ \sqrt{1 + \left(\frac{\lambda z}{\pi w_0^2}\right)^2} .
 \label{fwhm}
\ee
Note in this expression that the effect of increasing/decreasing $\lambda$ or $z$ is
equivalent; in both cases, the trend is analogous to the one discussed above for $z$ (in
relation to $z_R$), that is, the FW02M displays a hyperbolic dependence on the corresponding
parameter (nearly constant beginning and asymptotic linear increase, with
an intermediate boost in between).
If the input waist $w_0$ is considered instead, a rather different behavior is observed
on a projection screen at an output distance $z$ (and also for a fixed $\lambda$):
\begin{itemize}
	\item For $w_0 \ll \sqrt{\lambda z/\pi}$, i.e., for a relatively narrow input Gaussian beam,
	the width of the observed intensity distribution decreases with $w_0$, as
	\be
	{\rm FW02M} \approx \sqrt{2 \ln 5}\ \frac{\lambda z}{\pi w_0} ,
	\label{fraunregime}
	\ee
	which is exactly what is expected from a typical Fraunhofer regime.
		From this relation, it is seen that $z$ has been chosen in such a way that the
	intensity distribution is well inside the Fraunhofer regime.
	
	\item For $w_0 \gg \sqrt{\lambda z/\pi}$, i.e., for a relatively wide input Gaussian beam,
	the width of the intensity distribution increases linearly with $w_0$, as
	\be
	{\rm FW02M} \approx \sqrt{2 \ln 5}\ w_0 .
	\label{geomregime}
	\ee
	This result is in correspondence with typical geometrical optics regime,
	where the effective extension of the irradiance distribution at the output plane $z$ is
	proportional to its extension at the input plane (diffraction-free behavior).
\end{itemize}
In sum, a clear transition from a fully diffractive (Fraunhofer) regime to a seemingly
geometrical-optics one, each one with a very specific dependence on $w_0$, is observed.
Accordingly, at some point in between there should be a minimum in the
	width, denoting a turnover.
Getting back to Eq.~(\ref{fwhm}), for fixed $\lambda$ and $z$, we find that it has a
minimum when the input waist $w_0$ reaches the critical value
\be
w_c = \sqrt{\frac{\lambda z}{\pi}} .
\label{turnover}
\ee
Substituting this value into Eq.~(\ref{fwhm}) leads to the FW02M turnover value,
\be
{\rm FW02M}_t = 2 \sqrt{\ln 5}\ w_c = 2 \sqrt{\ln 5}\
\sqrt{\frac{\lambda z}{\pi}} \approx 1.4315 \sqrt{\lambda z}\ .
\ee

\begin{figure}[t]
	\centering
	\includegraphics[width=\textwidth]{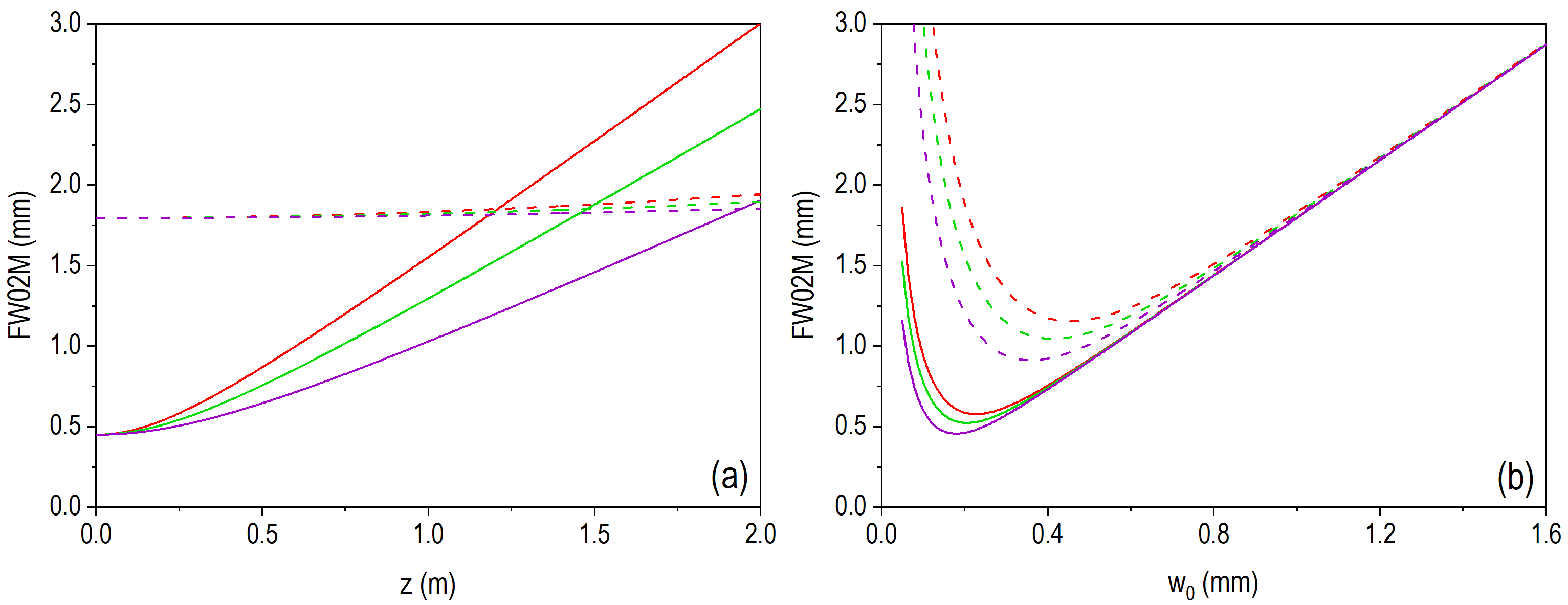}
	\caption{\label{Fig1}
		Full width at 20\% the maximum intensity (FW02M) for a diffracted Gaussian beam in terms
		of its distance between the input and output planes (a) and the size of its waist (b) for
		three different values of the wavelength: $\lambda = 650$~nm (red), $\lambda = 532$~nm
		(green), and $\lambda = 405$~nm (violet).
		To compare with, in panel (a) two input waist sizes are considered: $w_0 = 0.25$~mm (solid
		lines) and $w_0 = 1$~mm (dashed lines).
		Similarly, in panel (b) two input-output distances are considered: $z = 0.25$~m (solid lines)
		and $z = 1$~m (dashed lines).} 
\end{figure}

e numerical representation of Eq.~(\ref{fwhm}), describing the behavior of
	the FW02M, is displayed in Fig.~\ref{Fig1}.
	Although no experimental outcomes are reported here (all results shown arise from
	numerical calculations), realistic values have been chosen for the physical parameters
	involved, so that they can easily be reproduced even in a basic Optics laboratory.
Thus, the three wavelengths considered correspond to values of standard laser pointers,
$\lambda = 650$~nm, $\lambda = 532$~nm, and $\lambda = 405$~nm, with maximum output power
below 5~mW and uncertainties of the order of $\pm 10$~nm, but that basically cover the
visible spectral range.
The FW02M dependence on the distance between the input and output planes, $z$, as given
by Eq.~(\ref{fwhm}), is shown
in Fig.~\ref{Fig1}(a) for two values of the input waist, $w_0 = 0.25$~mm (solid lines) and
$w_0 = 1$~mm (dashed lines), and the above three wavelengths (each curve with the corresponding
color: red, green, and violet).
As it can be noticed, the wider the Gaussian beam at the input plane the lesser its spreading at the
output plane, which implies an also lesser difference among the results for
	the three wavelengths.
Furthermore, it is also clearly seen that, as the beam size decreases,
its three characteristic regimes, Huygens, Fresnel, and Fraunhofer, are more
	clearly distinguishable.
	For instance, for $w_0 = 0.25$~mm, the FW02M remains basically constant up to
$z \approx 0.05$~m, then it undergoes a nearly quadratic (hyperbolic) increase up to $z \approx 0.3$~m,
and finally it exhibits a linear increase regardless of the value of $z$.
For the wider beam, though, with $w_0 = 1$~mm, the linear regime is not reached within
	the range of 2~m here considered, but we can only observe the nearly constant width
	characterizing the Huygens regime for the three wavelengths.
Note that, if we consider an average wavelength $\bar{\lambda} = 529$~nm, the Rayleigh range
corresponding to $w_0 = 0.25$~mm is $\bar{z}_R = 0.37$~m, while for $w_0 = 1$~m we have
$\bar{z}_R = 5.94$~m, both is agreement with the values inferred from the numerical
results.
Yet, the gradual increase observed with $z$ does not allow us to set a clear distinction between
the Fraunhofer regime and the geometrical optics one.
The same happens if, instead of varying $z$, we had chosen to vary $\lambda$, according to
Eq.~(\ref{fwhm}).

In order to make apparent the transition from the Fraunhofer regime to
	a seemingly geometrical optics regime, let us now analyze the dependence of Eq.~(\ref{fwhm})
	on the value of the input waist $w_0$.
	The numerical representation of Eq.~(\ref{fwhm}) is shown in Fig.~\ref{Fig1}(b) for the same
	three wavelengths (same color code) and two different distances between the input and output
planes, $z$: $z = 0.25$~m (solid lines) and $z = 1$~m (dashed lines). 
The picture that emerges this time is quite different, with two well-defined trends for the
two distances and all wavelengths, one of them falling down with $w_0$, in agreement with
Eq.~(\ref{fraunregime}), and another one exhibiting a linear increase with $w_0$, as described by
Eq.~(\ref{geomregime}).
Furthermore, it is also clearly seen how an increase in the wavelength or in the distance
to the output plane produce an increase in the position of the transition input waist,
$w_c$, in agreement with the fact that decreasing any of these quantities (or both)
	enhances the wave (Fraunhofer diffraction) behavior of the beam, i.e., the FW02M will increase
	very quickly as a consequence of the fast dispersion of the beam at the Fraunhofer regime.
	On the contrary, this also implies that, to observe features typical of a geometrical optics
	regime, the width of the input beam should be pretty larger.
	Nonetheless, interestingly, in this latter case, in agreement with Eq.~(\ref{geomregime}),
	the FW02M becomes asymptotically independent of the wavelength $\lambda$ or the input-output
	distance $z$, which is the distinctive trait of a typically geometrical optics regime.
	These simple numerical examples thus show us how the same field (a Gaussian beam, in this
	case) can transition from a behavior typical of the wave optics to another characteristic of
	the geometrical optics by only gradually changing a single parameter.
	Furthermore, while in the latter regime the behavior is invariant with the wavelength or the
	observation distance, in agreement with the outcomes extracted from the traditional scalar
	theory of diffraction, in the fully wave (Fraunhofer) regime there is a clear dependence on
	those parameters, making the FW02M highly sensitive to small variations of them.
	This is clearly seen as $w_0$ decreases, getting closer to the critical value determining
	the transition threshold.
	Note that this critical value undergoes a displacement towards larger values of the input
	waist as either $\lambda$ or $z$ (or both) increase.
	For instance, with the values here considered, for the same wavelength, this displacement
doubles when the projection screen is moved from $z=0.25$~m to $z=1$~m.

Before concluding this section, we would like to briefly mention that the general shape
displayed by the graphs in Fig.~\ref{Fig1}(b), and particularly the presence of a turnover
that separates two physically different trends or behaviors, to some extent resembles a
result found in quantum gravity, for micro-black holes, in the context of the so-called
generalized uncertainty principle \cite{scardigli:PhysLettB:1999}.
Within this scenario, the quantity that plays the role of the critical beam width $w_0$
is the Planck energy, observing a similar trend between a given space region of a certain
width and the energy that it contains.
If quantum fluctuations are important, then there is an inverse relation between the two
quantities; if classical gravitation is relevant, then they are proportional.
The critical length at which these two behaviors coincide is precisely the Planck length
(for energy uncertainties of the order of the Planck energy), at which a micro black hole
could originate.


\section{Diffraction by a long rectangular slit}
\label{sec3}


\subsection{Theoretical analysis}
\label{sec31}

Let us now consider the diffraction through a very long rectangular slit, with the $x$-axis
in the direction of the shorter opening, with a width $a$, and the $y$-axis
	along the longer one, with a width $b \gg a$.
Due to translation symmetry along the $y$-direction, the problem can be solved and
explained, in a good approximation, within the $XZ$-plane, where $x$ will denote the
transverse coordinate and $z$ the longitudinal (propagation) one.
Appealing again to paraxial conditions, the electric field amplitude, solution of the
(paraxial) Helmholtz equation, can be recast \cite{bornwolf-bk} as
\be
E(x,z) \propto E_0 \int_{-a/2}^{a/2} e^{i\pi (x-x')^2/\lambda z} dx' ,
\label{Efield1}
\ee
in the case of an incident field with constant amplitude within the opening determined
by the slit, from $-a/2$ to $a/2$, and where prefactors and global phase factors are
neglected for simplicity, but without any loss of generality (a detailed derivation of the
general paraxial solution can be found in \cite{sanz:AOP:2015} in the context of matter
waves and the Schr\"odinger equation).
If the output plane $z$ is fixed, Eq.~(\ref{Efield1}) can be rewritten as
	\be
	E(x,z) \propto E_0 \int_{-a/2}^{a/2} e^{i(x-x')^2/w_c^2} dx' ,
	\label{Efield2}
	\ee
	where $w_c$ coincides with the value given by Eq.~(\ref{turnover}).
	Nonetheless, note that this parameter now describes a general distance
	that can be used to compare with, i.e., it is not the waist of a Gaussian beam.

As in Sec.~\ref{sec2}, here we also consider the relative intensity distribution,
\be
I_{\rm rel}(x,z) = \frac{I^2(x,z)}{I^2(0,z)} \propto
\left\vert \int_{-a/2}^{a/2} e^{i (x-x')^2/w_c^2} dx' \right\vert^2 ,
\label{geninteg}
\ee
where $I(x,z) \propto |E(x,z)|^2$, $x'$ and $x$ describe positions on the input and output
plane, respectively (which correspond to the planes where the slit and the projection
screen are accommodated). 
In principle, the next step in the analysis would consist, also as in Sec.~\ref{sec2}, in
investigating the behavior of (\ref{geninteg}) and determining the different propagation
regimes.
However, this task is not analytical for intermediate stages at the Fresnel
regime, which requires the use of numerical techniques.
Yet, the integral is simple enough to allow us extracting some physical
	insight leaving aside further
calculations (even though they are eventually necessary to determine and
understand the full trend).
Thus, consider the integral
\be
\mathcal{I}(x,z) = \int_{-a/2}^{a/2} e^{i({x'}^2 - 2 x x')/w_c^2} dx' ,
\label{integral}
\ee
where the factor $e^{ix^2/w_c^2}$ is disregarded, because it is eventually
	suppressed in (\ref{geninteg}), and hence it has no physical relevance at all here.
In order to make apparent the three regimes described in Sec.~\ref{sec2}, the question to
be addressed now is whether, for a given $z$ (or, in general, a given value of the typical
length scale $w_c$), the phase factors ${x'}^2$ and $2xx'$ are relevant,
	and, in the affirmative case, which one of them is the leading one.
When proceeding in this way, the following scenarios readily arise for a given wavelength of
interest:
\begin{itemize}
	\item If $z \approx 0$, the integrand becomes a very rapidly oscillatory function.
	On average, one could then assume that only when this function is evaluated over the
	actual point $x$ [i.e., $x' \approx x$ in the integrand of Eq.~(\ref{geninteg})], there is
	a non-vanishing contribution to the integral; otherwise, it is rapidly
		vanishing.
	If this is extended to the full integration range, in a good approximation the result of
	the integral (\ref{integral}) is a constant, namely, $\mathcal{I}(x,z) \approx a$.
	This corresponds to a geometrical regime, where there is a nearly constant irradiance
	in front of the slit, surrounded on either side by a region of geometrical shadow.
	This thus corresponds to the Huygens regime, ruled out by geometrical optics, which holds
	for either small distances between input and output planes, but also, if $z$ is fixed,
	for negligible wavelengths.
	
	\item For longer but small enough values of $z$, so that the values $x$ of interest
	(where the intensity is important) are still within the $(-a/2,a/2)$ interval or nearby,
	the trend is similar, although the phase terms start playing a role, particularly, the
	${x'}^2$ term, which is typically smaller and hence leads to important oscillations.
	This quadratic dependence on the slit coordinate is typical of the Fresnel regime and
	physically manifests as the appearance of rather prominent oscillations in the intensity
	distribution, with some leaks towards the 
	geometrical shadow region, even though it still mainly concentrates within the area covered 
	by the slit.
	Note that, if the second phase term is neglected in Eq.~(\ref{integral}), the Fresnel
	integrals readily appear, which leads to the usual methods in wave optics developed to
	determine analytically the on-axis intensity (Fresnel zones and the Cornu spiral).
	
	\item Finally, at rather long values of $z$ , far beyond the slit input plane, it is the
	second phase term the one that becomes prominent, since $x$ may acquire values beyond the
	interval $(-a/2,a/2)$.
	In this case, the integral (\ref{integral}) has a simple analytical solution:
	\be
	\mathcal{I}(x,z) \sim \int_{-a/2}^{a/2} e^{-2 i x x'/w_c^2} dx'
	\propto {\rm sinc} \left( \frac{\pi a x}{\lambda z} \right) .
	\label{integralsinc}
	\ee
	With this, the intensity (\ref{geninteg}) becomes
	\be
	I_{\rm rel}(x,z) = {\rm sinc}^2 \left( \frac{\pi a x}{\lambda z} \right) ,
	\label{fraunhofer}
	\ee
	which is the typical intensity distribution generated by a single slit in the Fraunhofer
	regime.
		The profile displayed by this intensity distribution at the output plane, regardless of the
		value of $z$, only depends on the ratio $x/z$, which determines the observation angle, $\theta \approx x/z$, in paraxial approximation.
\end{itemize}
We thus see in a simple and intuitive manner that, effectively, the three regimes are a
general trait of any diffraction process, independently of the shape of the initially
diffracted field (the field at the input plane).

Let us know get back to the question of determining the width of the intensity distribution
at a fixed output plane $z$, also measured in terms of the FW02M.
From the above discussion, and following the reasoning given in \cite{panuski:PhysTeacher:2016},
a suitable guess in the geometrical optics regime for the FW02M is
\be
{\rm FW02M} \propto a ,
\label{geomfwhm}
\ee
assuming that Fresnel diffraction features are negligible, at least, at first approximation.
Of course, some deviations should be expected, but in the limit a very large slit width,
the approximation should be good enough.
On the contrary, for tiny slit widths, one should expect Fraunhofer diffraction to be
dominant and a decreasing trend for FW02M with increasing $a$, as it is inferred from
Eq.~(\ref{fraunhofer}).
In order to determine an analytical expression for the FW02M in this case, let us consider
the sine approximation formula developed by the VIIth-century Indian astronomer and
mathematician Bhaskara I \cite{shirali:MathMag:2011},
\be
\sin u \approx \frac{16 u (\pi - u)}{5\pi^2 - 4u(\pi - u)} ,
\ee
where $u$ is measured in radians ($0 \le u \le \pi$).
From this formula, it follows that the sinc-function can be approximated as
\be
{\rm sinc}\ \! u \approx \frac{16 (\pi - u)}{5\pi^2 - 4u(\pi - u)} .
\ee
Here, we have $u = \pi a x /\lambda z$.
If we search for the value of $u$, such that ${\rm sinc}^2\ \! u = \beta^2$, we find
\be
u = \frac{1}{2} \left( \pi - \frac{4}{\beta} + \frac{2}{\beta}
\sqrt{4 + 2\pi \beta - \pi^2 \beta^2} \right) .
\ee
After substituting $\beta = \sqrt{0.2} = 1/\sqrt{5}$ into this latter expression, we find
\be
u = \frac{\pi}{2} - 2\sqrt{5}
+ \sqrt{20 - \pi^2 + 2\pi\sqrt{5}} \approx 2.01598
\ee
(the negative root is neglected, because of the domain of definition for $u$ indicated
above).
Therefore, the expression of the FW02M in the Fraunhofer regime reads as
\be
{\rm FW02M} \approx \frac{2 u \lambda z}{\pi a} \approx 1.28341\ \frac{\lambda z}{a} ,
\label{fraunfwhm}
\ee
which shows the expected dependence on the inverse of the slit width $a$.
Since there is not an analytical expression for the FW02M when the full
	range of slit widths is covered, as it happens with Gaussian slits, to determine in an
	approximate manner the critical width $a_c$ we assume that, for this value,
	Eqs.~(\ref{geomfwhm}) and (\ref{fraunfwhm}) must be equal.
This renders a critical slit-width value
\be
a_c \approx 1.13288\ \sqrt{\lambda z} \approx 2.00797\ \sqrt{\frac{\lambda z}{\pi}} .
\label{acritical}
\ee
If this value is substituted into Eq.~(\ref{fraunfwhm}), we obtain the approximated value
for the turnover FW02M,
\be
{\rm FW02M}_t \approx 1.13287\ \sqrt{\lambda z}
\approx 2.007966\ \sqrt{\frac{\lambda z}{\pi}} ,
\label{FW02Mcrit}
\ee
which is about twice the turnover value found for the Gaussian beam.

Of course, the above conclusions are just reasonable analytical conjectures, which do not
	provide further details about the transition that we are studying here, and that must
	be validated in some way.
	Here, this will be done by numerically determining the FW02M.
	In particular, in order to get a deeper insight, next two intertwined routes
	are considered, namely, a numerical analysis, which renders some light on the trends
	analytically found (as it was done in the case of Gaussian slits), and then a (numerical)
	comparison with the experiment, with the purpose to verify the validity of such numerical
	analysis.


\subsection{Numerical analysis}
\label{sec32}

\begin{figure}[t]
	\centering
	\includegraphics[width=\textwidth]{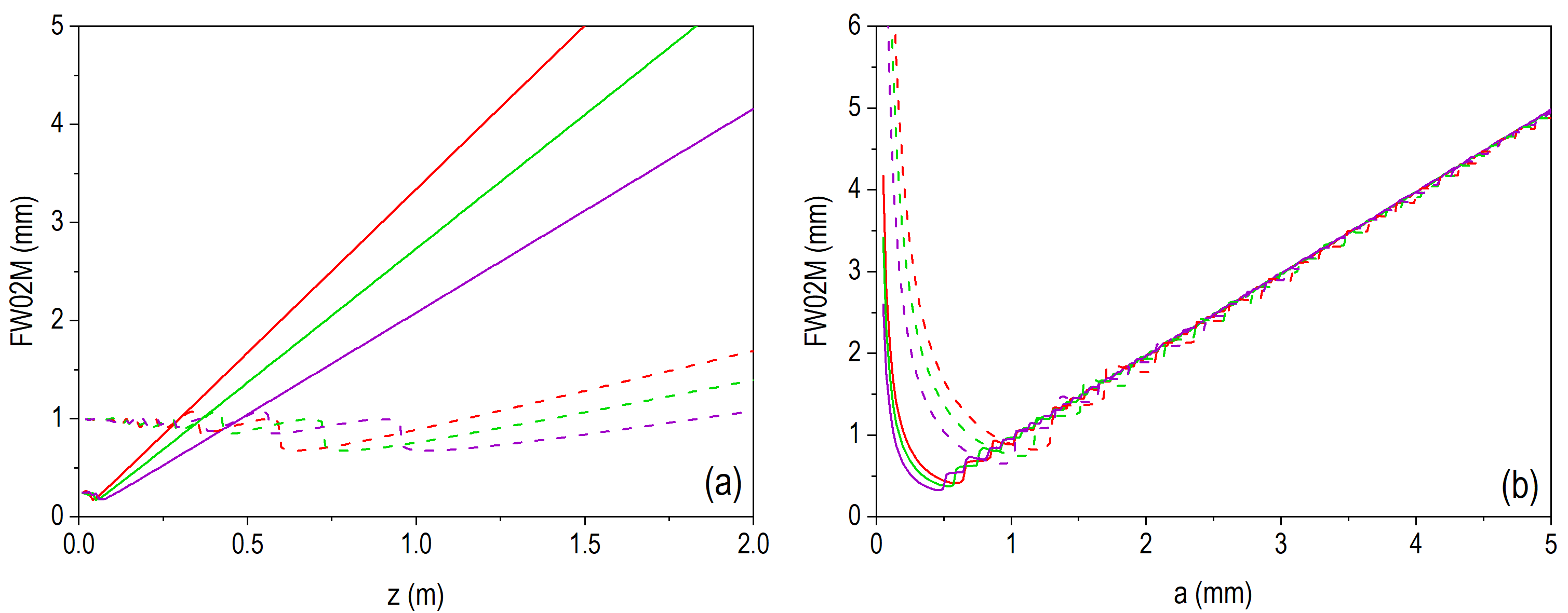}
	\caption{\label{Fig2}
		Full width at 20\% the maximum intensity (FW02M) for the diffraction of a monochromatic
		plane waves incident on a very long rectangular slit in terms of the distance between the
		input and output planes $z$ (a) and the slit width $a$ (b) for three different values of
		the wavelength: $\lambda = 650$~nm (red), $\lambda = 532$~nm (green) and
		$\lambda = 405$~nm (violet).
		To compare with, in panel (a) two slit widths are considered: $a = 0.25$~mm (solid lines)
		and $a = 1$~mm (dashed lines).
		Similarly, in panel (b) two distances are considered: $z = 0.25$~m (solid lines) and
		$z = 1$~m (dashed lines).} 
\end{figure}

As in Sec.~\ref{sec2}, in Fig.~\ref{Fig2} we show the dependence of the FW02M on the
distance between the input and output planes (a), and on the slit width $a$ (b), which is
the analog to the Gaussian beam waist $w_0$.
In each case, the same three wavelengths have been considered.
More specifically, given the lack of analyticity of the FW02M along the
full range of slit widths considered, it has to be numerically computed.
Thus, to carry out the integral (\ref{integral}) that the intensity (\ref{geninteg}) is
based on, a simple on-purpose Fortran code was programmed (based on the trapezoid rule,
which suffices by far in this case), considering both a large box (in order to
accommodate wide intensity distributions in both limits) and a large number of sampling
points (to resolve the fastest oscillations in the Fresnel regime).
Fixing the value of two parameters, the code automatically runs over a large set of values
of the remaining parameter, thus rendering a smooth graph for the FW02M, analogous to the
the graphs obtained from the analytical expressions for the Gaussian slit.
In particular, the code computes the relative intensity distribution.
(\ref{geninteg}), normalizes its maximum to unity, and then finds the $x$-positions ($x_1$
and $x_2$, symmetrically distributed around $x=0$) at which the latter equals 0.2 in order
to determine the FW02M.

Proceeding in that way, we have thus obtained the results shown in Fig.~\ref{Fig2}(a) for two
slit widths: $a = 0.25$~mm (solid lines) and $a = 1$~mm (dashed lines).
Consider the case for the smallest slit width.
Unlike the smooth dependence exhibited by the Gaussian beam, it is observed that there is a
very short Huygens regime, followed by a saw-tooth structure characterizing the intermediate
Fresnel regime; immediately afterwards, the Fraunhofer regime starts, which is clearly
distinguished by its distinctive linear trend.
If the slit width increases, the same behavior is observed, although the distance $z$ at
which the Fraunhofer regime is reached increases very rapidly.
Note that, on average, an increase from 0.25~mm to 1~mm in $a$ implies an increase from $\sim 0.05$~mm to $\sim 1$~mm.
Although the near field structure is more complex than the same region for the Gaussian
beam, we find again the same lack: it is difficult to observe the transition that we are
investigating, because we cannot see a clear turnover.

To overcome that problem, we proceed as before and compute the FW02M against the slit width,
which is shown in Fig.~\ref{Fig2}(b) for the same three wavelengths and two values of the
distance between the input and output planes: $z = 0.25$~m (solid lines) and $z = 1$~m
(dashed lines).
Thus, the numerical simulations render two important general trends.
First, an initial falloff with increasing (but small values of) $a$, which is in compliance with
the behavior described by Eq.~(\ref{fraunfwhm}), i.e., with the inverse of the slit width.
Second, an incipient overall linear trend is observed for large values of $a$, which is in
correspondence with the guess (\ref{geomfwhm}).
Furthermore, we also find that there is a turnover denoted by the presence of a minimum in the
representation, as in the case of the Gaussian beam [see Fig.~\ref{Fig1}(b)], which
approximately coincides with the values estimated above for $a_c$ and ${\rm FW02M}_t$.
For instance, if we consider the average wavelength $\bar{\lambda} = 529$~nm, the turnover
is obtained at $a_c \approx 0.4$~mm for $z = 0.25$~m, and at $a_c \approx 0.8$ for
$z = 1$~m, which coincides with what we observe in Fig.~\ref{Fig2}(b).
In either case, though, these values are larger than for a Gaussian beam, because the expansion of
``top-hat'' type diffracted beams is typically faster, which requires larger openings to observe
slower diffractive rates.
Nonetheless, it seems that the preliminary analytical treatment, based on a series of reasonable
work hypotheses, is not that bad.
However, what such an analytical treatment does not describe is the presence of the staircase structure
that starts at the transition between the end of the Fraunhofer regime and the beginning of the Fresnel
one, and somehow influences the turnover region, avoiding us to unambiguously identify a minimum
(note that the minimum observable now is affected by such a staircase structure).
Yet the extrapolation of a linear regression at large values of $a$, where the steps become smaller
and smaller, allows us to determine the turnover from the intersection between this linear fitting
and the Fraunhofer falloffs, which render basically the estimated values mentioned before.

\begin{figure}[t]
	\centering
	\includegraphics[width=\textwidth]{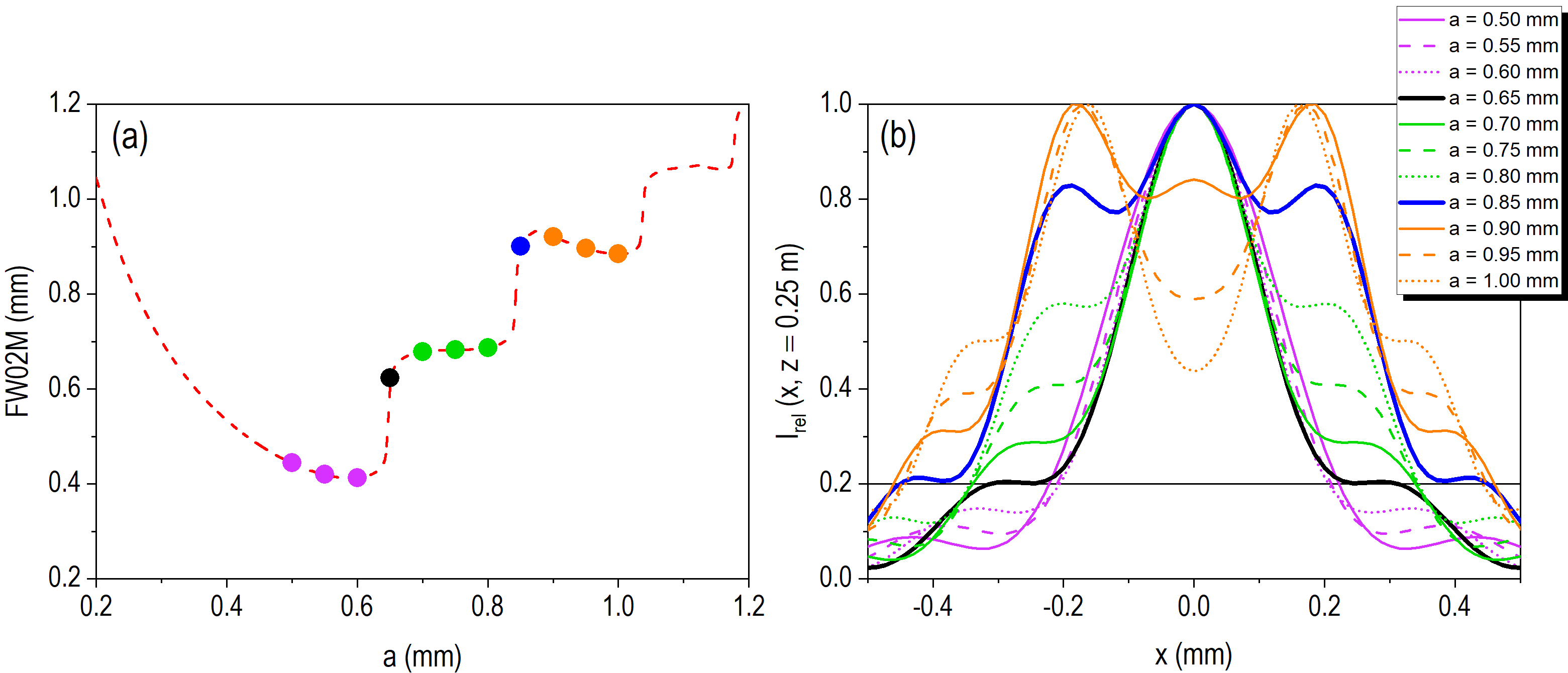}
	\caption{\label{Fig3}
		(a) Detail of the full width at 20\% the maximum intensity (FW02M) displayed in
		Fig.~\ref{Fig2} for $\lambda = 650$~nm and $z = 0.25$~m.
		Solid circles denote the sampling values considered to analyze the staircase structure
		displayed by the FW02M around the turnover region.
		(b) Relative intensity distributions (at $z=0.25$~m) associated with the selected values
		of slit widths shown in panel (a).
		Same color line (but different type) is used for values on the same step; thicker lines
		correspond to values at the jump.}
\end{figure}

The staircase structure of Fig.~\ref{Fig2}(b) is connected to the systematic method employed by our
numerical code, which determines the precise value where the relative intensity reaches the 20\% of
its maximum value.
In order to determine the origin of this distinctive structure, absent in the case of Gaussian beams, let us focus on an
enlargement of Fig.~\ref{Fig2}(b) around the turnover region, displayed in Fig.~\ref{Fig3}(a).
In the latter figure, the turnover region is seen for a wavelength $\lambda = 650$~nm and a distance
$z = 0.25$~m.
The numerical simulation is denoted with the black solid line.
Some sampling values associated with different slit widths are also shown (red and blue solid circles),
which cover three step levels and will be used to elucidate the origin of
	these steps (in particular,
the blue circles are markers related to the border between one step and the next one).
The relative intensities associated with each marker are shown in
	Fig.~\ref{Fig3}(b) with different colors, for slit widths ranging from 0.5~mm to 1~mm, in
tiny increments of 0.05~mm.
As it can be noticed, because the turnover region is fully immersed in the Fresnel regime
(actually, for moderate values of $a$, we should talk about the intermediate regime between
the Fraunhofer and Fresnel ones \cite{elmore-heald-bk:1969}), such small increments may
induce important changes, as it is seen near the sudden jumps from one step to the next one.
What happens is that the intensity is characterized by marginal oscillating ``wings'',
as seen in Fig.~\ref{Fig3}(b), which start developing as the Fraunhofer regime blurs, and
diffraction minima no longer vanish, particularly those adjacent to the principal maximum.
The general behavior of these minima is that they start increasing, progressively elevating
with them the secondary maxima and generating a highly wavy intensity distribution as $a$
increases.
Each time that one of these secondary maxima reaches the 20\% intensity threshold, we
observe a sudden increase in the corresponding FW02M.
For example, in Fig.~\ref{Fig3}(b), this is seen to happen for $a = 0.65$~mm (green thick
solid line) and $a = 0.85$~mm (black thick solid line); in the first case, the jump is
associated with the disappearance of the first adjacent secondary maxima, while in the
latter case the jump is connected to the second one.
Note that, eventually, the intensity pattern displays the well-known profile of a nearly flat
distribution, approximately along the same extension covered by the slit, modulated by a
series of small-amplitude and rapidly-varying oscillations both at the top and also on the
sides (in the regions of geometrical shadow).
This is the typical behavior that is provided in Optics textbooks to illustrate
	the Fresnel regime \cite{hecht-bk:2002,bornwolf-bk,elmore-heald-bk:1969}, although, as
	seen in Fig.~\ref{Fig2}(b), it quickly approaches a nearly linear trend as $a$ increases
	well above $a_c$.


\subsection{Comparison with the experiment}
\label{sec33}

\begin{figure}[t]
	\centering
	\includegraphics[width=0.8\textwidth]{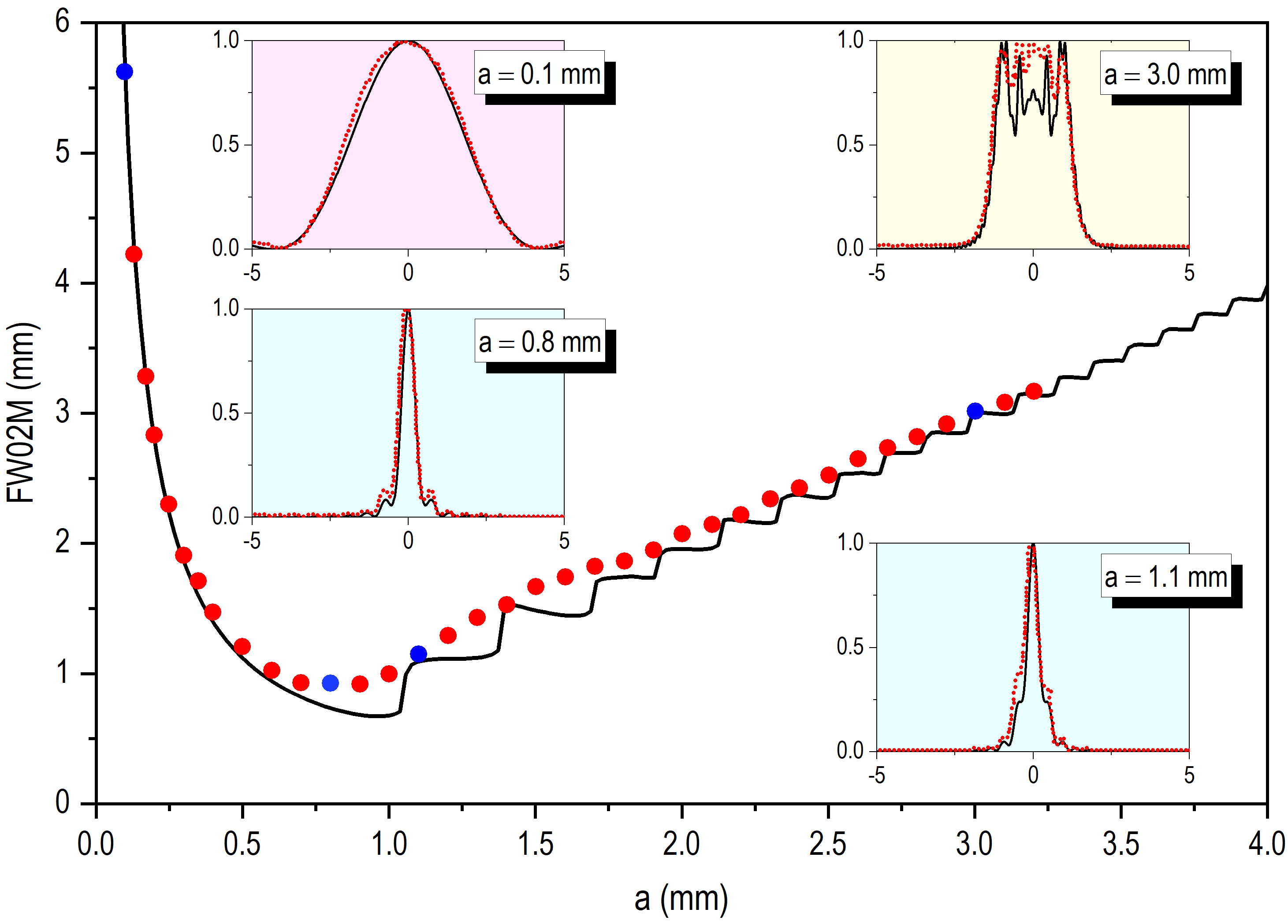}
	\caption{\label{Fig4} Comparison with the experimental data reported in
		\cite{panuski:PhysTeacher:2016}, with $\lambda_{\rm exp} = 660$~nm and
		$z_{\rm exp} = 0.656$~m.
		In the main panel, the experimental data are denoted by solid circles, while the
		theoretical simulation following Eq.~(\ref{geninteg}) is described by the black solid
		line.
		In the insets, relative intensity for four different values of the slit width $a$:
		$a = 0.1$~mm, $a = 0.8$~mm, $a = 1.1$~mm, and $a = 3.0$~mm.
		Again here the red solid circles refer to experimental data, while the solid line
		indicates the intensity rendered by the theoretical model, Eq.~(\ref{geninteg}).
		These four particular cases are denoted in the main panel with blue solid circles and
		have been picked up in different regimes, two in the limiting cases (Fraunhofer and
		geometrical optics regimes) and two in the turnover region.}
\end{figure}

In order to test the validity of the approach that we have followed here,
let us now compare with the experimental data reported in \cite{panuski:PhysTeacher:2016}.
	An alternative theoretical analysis has been previously reported in
\cite{davidovic:PhysTeacher:2019} considering the variation of the intensity distribution
in terms of the Fresnel number (zones), $FN = a^2/4\lambda z$.
In Fig.~\ref{Fig4} we show the numerically computed FW02M as a function of the slit width, which has been obtained by using the same raw values for $\lambda$ and $z$
	considered in the experiment reported in \cite{panuski:PhysTeacher:2016}, namely
	$\lambda_{\rm exp} = 660$~nm and $z_{\rm exp} = 0.656$~m, without any extra fitting
	and/or numerical treatment.
	As it is seen, the theoretical (numerical) representation exhibits the typical staircase structure beyond the turnover region above described,
which gradually approaches a nearly linear tail for large $a$.
Comparing the experimental data (solid circles), directly extracted from  \cite{panuski:PhysTeacher:2016} and inserted in the representation (also without any
	extra treatment), with the theoretical (numerical) results, it is found that there is a
	good agreement, in particular, concerning the overall trend.
	The staircase structure, though, is not observed in the experiment.
After closely inspecting and analyzing the experimental data, particularly observing the
slight discrepancies between the theoretical and experimental intensity distributions (some
of which are represented in the insets for the slit widths indicated), we come to the
conclusion that, in order to observe the staircase structure, the quality of the data
recording procedure requires some further refinement, for small fluctuations around the
sudden jumps will suppress the effect.
Nonetheless, getting back to the experimental data shown in Fig.~\ref{Fig4}, we find
that the turnover region is around the critical value that we directly obtain from
Eq.~(\ref{acritical}) for $\lambda_{\rm exp}$ and $z_{\rm exp}$, namely,
$a_c \approx 0.74$~mm, which is closer to the experimental turnover than the former
estimate, $a_c \approx 0.931$~mm, provided in \cite{panuski:PhysTeacher:2016}.


\section{Final remarks}
\label{sec4}

In this work, an analysis of the transition from the Fraunhofer diffraction regime to
the geometrical optics limit has been carried out with the purpose to better understand
the meeting point between wave optics and geometrical optics, and hence
	to get a deeper insight into the physical origin and consequences of the latter,
	alternative to wavelength-based considerations.
Motivated by the experiments carried out by Panuski and Mungan \cite{panuski:PhysTeacher:2016},
which show how this transition takes place in a nice manner, beyond such wavelength-based
considerations, here we have tackled the issue by first investigating the behavior of
Gaussian beams and then the case of single slit diffraction.
	In both cases it has been observed how, by means of analytical treatments, whenever the
	width of the intensity profile is analyzed in terms of the effective slit width (the beam
	waist or the actual slit width), a turnover critical value for this width readily arises, which depends on two parameters, namely,
the wavelength $\lambda$ of the monochromatic source considered, and the distance $z$
between the input (slit) and output (projection screen) planes.
Analytical expressions have been found for the critical values in the two cases considered,
	both being proportional to $\sqrt{\lambda z}$.
Accordingly, it is seen that the rapid falloff with the inverse of the slit width, which
characterizes the width of the intensity distribution in the Fraunhofer regime, gives rise
to a linearly increasing width of such a distribution beyond the critical value, in
correspondence with the expectations from a geometrical optics point of view.

Theoretical (numerical) results have also been compared to experimental
	data extracted from \cite{panuski:PhysTeacher:2016}, finding a good agreement and, in
	particular, a systematic way to determine the value of the turnover condition from
	a theoretical model.
	Actually, the numerical results have shown evidence that, while the Fraunhofer falloff is
	clearly seen, the transition towards the geometrical regime goes through a staircase
	structure with shorter and shorter steps, starting within the turnover region.
	This structure is directly related to the gradual suppression of the typical Fraunhofer
	diffraction pattern, where each step in the staircase corresponds to a sort of quantized
	increase of the FW02M coming from the adjacent secondary maxima.
	Although the experimental data reported in \cite{panuski:PhysTeacher:2016} do not allow
	to observe this structure, they fit pretty well the asymptotic behavior towards the
	geometrical regime.

To conclude, it is worth mentioning that the treatments and findings account for here
can be straightforwardly extended to matter waves.
As it was mentioned above, in Sec.~\ref{sec2}, the isomorphism between the paraxial
Helmholtz equation and the time-dependent Schr\"odinger equation enables a direct
translation of the findings reported here to matter waves, which, in principle, should
be observable with the appropriate experimental conditions.
Since fundamental aspects of interference have been explored with electrons
\cite{batelaan:NJP:2013,batelaan:NJP:2018} and large molecular complexes
\cite{arndt:Nature:1999,arndt:NatCommun:2011,arndt:NNanotech:2012}, these
systems could also be used to investigate the behaviors here observed.
Another interesting extension of the present study is that
of the so-called fractional Schr\"odinger equation \cite{longhi:OptLett:2015,zhang:PRL:2015,zhang:LaserPhotRev:2016},
which also applies to both massive quantum particles and light (where it is
used in the form of a fractional paraxial Helmholtz equation), and is governed
by a fractional spatial derivative.
In this latter regard, the action of a fractional Laplacian over the diffracted beam
is expected to render novel limits in both the Fraunhofer diffraction regime and the
geometrical optics one.


\section*{References}


\providecommand{\newblock}{}

\end{document}